\documentclass[manuscript]{acmart}
\usepackage{booktabs} 
\usepackage{multirow,tabularx}
\usepackage{threeparttable}
\usepackage{caption}
\usepackage[justification=centering]{subcaption}

\usepackage{color}
\graphicspath{{./Figures/}}
\DeclareGraphicsExtensions{.pdf,.jpeg,.jpg,.png,.tiff}

\newcommand{\ldq}[1]{{\color{black}#1}}

\renewcommand\footnotetextcopyrightpermission[1]{} %

\begin{document}

\copyrightyear{2017}
\acmYear{2017}
\setcopyright{acmcopyright}
\acmConference{MM'17}{October 23--27, 2017}{Mountain View, CA,
USA}\acmPrice{15.00}\acmDOI{10.1145/3123266.3123322}
\acmISBN{978-1-4503-4906-2/17/10}

\title{Exploiting High-Level Semantics for No-Reference Image Quality Assessment of Realistic Blur Images}

\author{Dingquan Li$^{1,3}$, Tingting Jiang$^{1,2}$, Ming Jiang$^{1,3}$}
\orcid{0000-0002-5549-9027}
\affiliation{%
  \institution{$^1$National Engineering Lab for Video Technology, Cooperative Medianet Innovation Center; $^2$School of EECS; $^3$LMAM, School of Mathematical Sciences \& BICMR, Peking University, Beijing 100871, China}
}
\email{{dingquanli,ttjiang,ming-jiang}@pku.edu.cn}

\renewcommand{\shortauthors}{D. Li et al.}

\begin{abstract}
To guarantee a satisfying Quality of Experience (QoE) for consumers, it is required to measure image quality efficiently and reliably. The neglect of the high-level semantic information may result in predicting a clear blue sky as bad quality, which is inconsistent with human perception. Therefore, in this paper, we tackle this problem by exploiting the high-level semantics and propose a novel no-reference image quality assessment method for realistic blur images. Firstly, the whole image is divided into multiple overlapping patches. Secondly, each patch is represented by the high-level feature extracted from the pre-trained deep convolutional neural network model. Thirdly, three different kinds of statistical structures are adopted to aggregate the information from different patches, which mainly contain some common statistics (\textit{i.e.}, the mean\&standard deviation, quantiles and moments). Finally, the aggregated features are fed into a linear regression model to predict the image quality. Experiments show that, compared with low-level features, high-level features indeed play a more critical role in resolving the aforementioned challenging problem for quality estimation. Besides, the proposed method significantly outperforms the state-of-the-art methods on two realistic blur image databases and achieves comparable performance on two synthetic blur image databases.
\end{abstract}

%

\keywords{Image Quality Assessment; Semantic; Realistic Blur; No-Reference Quality Metric; Statistical Aggregation Structure}

\maketitle

\section{Introduction}
\label{sec:introduction}
In the era of big data, images have become the primary carrier of information in human's daily life. Before ultimately received by a human observer, digital images may suffer from a variety of distortions. Quality of Experience (QoE), whose goal is to provide a satisfying end-user experience, has drawn increasing attention. To reach this goal, a critical precondition is to conduct image quality assessment (IQA). The most reliable way to assess image quality is subjective ratings, but it is often cumbersome, expensive and difficult to carry out in reality. Thus, objective IQA methods that can automatically predict image quality efficiently and effectively are needed. Objective IQA can be categorized into full-reference IQA (FR-IQA)~\cite{wang2004image,you2011modeling}, reduced-reference IQA (RR-IQA)~\cite{zhai2012psychovisual,qi2015reduced} and no-reference IQA (NR-IQA)~\cite{gu2015using,zhang2014objective}. Due to the unavailability of the reference image in most practical applications, NR-IQA is preferable but also more challenging.

In this paper, we focus on NR-IQA of realistic blur images. Blur is often induced by following reasons: (1) out-of-focus, (2) relative motion between the camera and the objects (object motion \& camera shake), (3) non-ideal imaging systems (\textit{e.g.}, lens aberration), (4) atmospheric turbulence, and (5) image post-processing steps (such as compression and denoising)~\cite{ferzli2009no,narvekar2011no,hassen2013image}. Except the blur in Bokeh to strengthen the photo's expressiveness, it is a definite fact that unintentional blur impairs image quality.

\begin{figure}[!htb]	
	\centering
	\begin{subfigure}[t]{2.8in}
		\centering
		\includegraphics[width=2.8in]{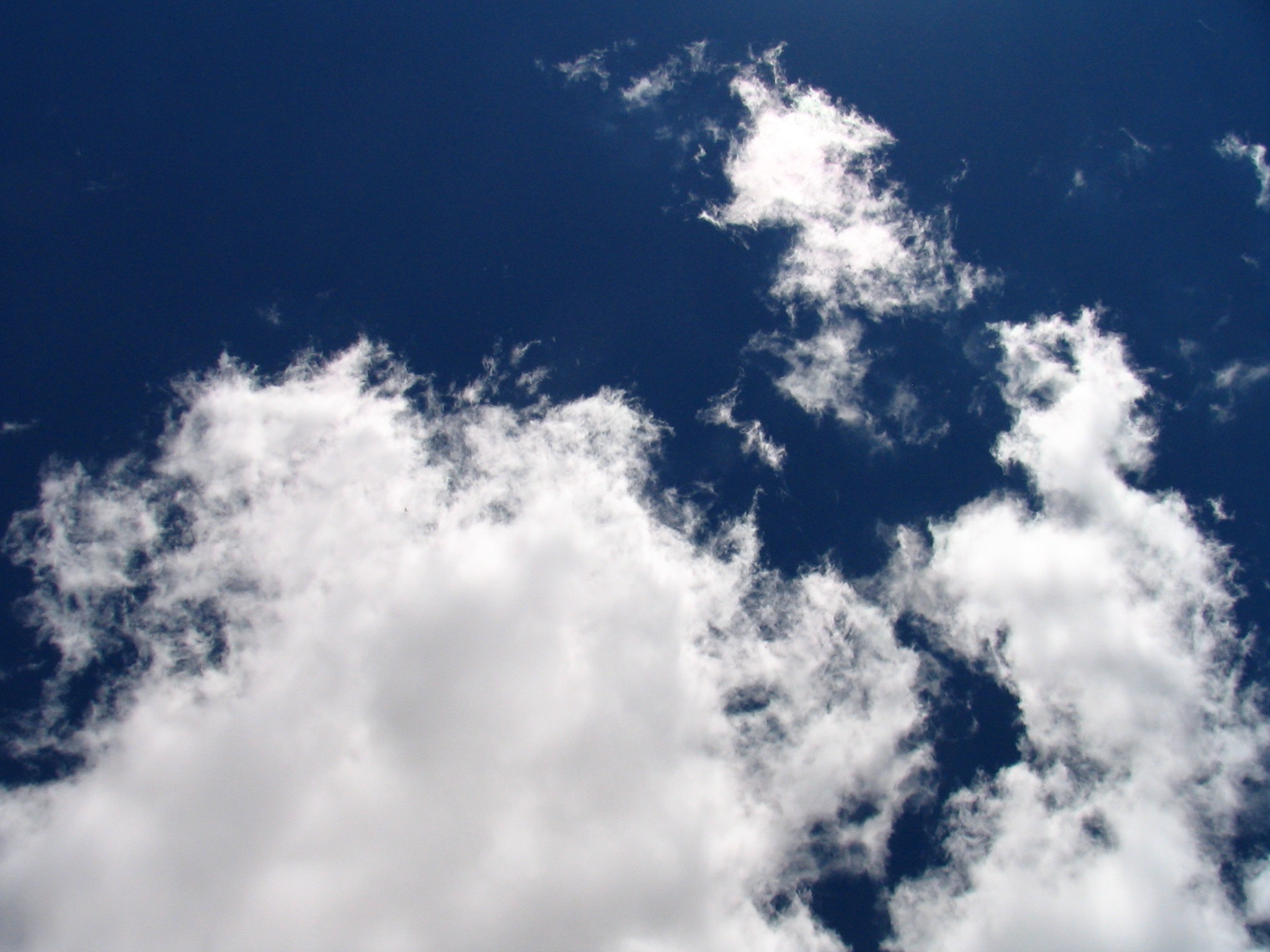}
		\caption{MOS=4.0637}
	\end{subfigure}
	\begin{subfigure}[t]{2.8in}
		\centering
		\includegraphics[width=2.8in]{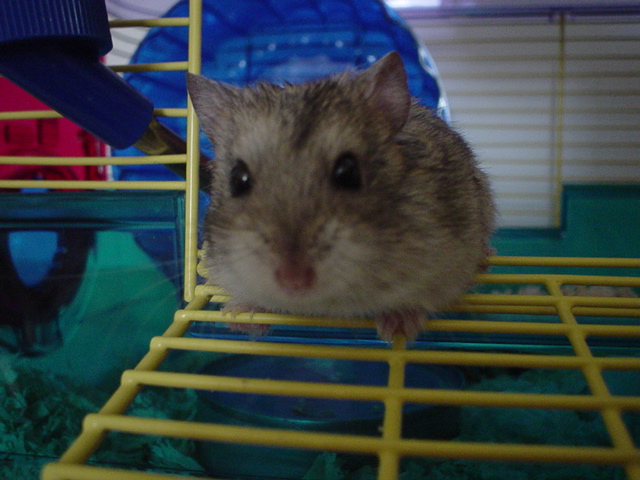}
		\caption{MOS=2.3413}
	\end{subfigure}
	\caption{The two images are from BID~\cite{ciancio2011no}, and larger MOS indicates better subjective image quality. The three traditional methods (MDWE~\cite{marziliano2002no}, FISH~\cite{vu2012fast}, LPC~\cite{hassen2013image}) predict that (a) is worse than (b). Our method predicts that (a) is better than (b), which is consistent with subjective ratings.}\label{fig:Success}
\end{figure}
\vspace{-2mm}
Traditional NR-IQA methods of blur images are mainly based on the assumptions that blur leads to the spread of edges (\textit{e.g.}, MDWE~\cite{marziliano2002no}), the reduction of high-frequency energy (\textit{e.g.}, FISH~\cite{vu2012fast}) or the loss of local phase coherence (\textit{e.g.}, LPC~\cite{hassen2013image}). However, these methods neglect the high-level semantic information, and can distinguish neither between intrinsic flat regions and blurry regions, nor between structures with and without blurring. As a result, it is shown in Figure~\ref{fig:Success} that they predict the quality of a clear sky being worse than the quality of a blurry mouse, which is not consistent with human perception.

In this work, we tackle the problem by exploiting the high-level semantic features extracted from the pre-trained deep convolutional neural network (DCNN) models. First of all, since the pre-trained DCNN models (\textit{e.g.}, AlexNet~\cite{krizhevsky2012imagenet}) require a fixed input size, we need to determine \textbf{how to represent an image}. We compare four different image representations, and find that the multi-patch representation significantly better than the others. Secondly, we need to decide \textbf{which pre-trained DCNN model and which layer to extract image features.} We first explore the effectiveness of features extracted from different layers in a same pre-trained DCNN model, and find out high-level features from the top third or second layer more effective in realistic blur image assessment. Then we investigate the impact of different pre-trained models, and find out the one using residual learning (\textit{i.e.},ResNet-50~\cite{he2016deep}) more suitable for NR-IQA of realistic blur images. Thirdly, as a result of the multi-patch representation, we derive a set of features for an image. So another question arises: \textbf{how to aggregate a set of extracted features?} One simple way is to use the mean feature vector to represent the feature set. However, it will lose important information (\textit{e.g.}, the standard deviation in each dimension) of the feature set. So we propose three different statistical structures for feature aggregation, namely, mean\&std aggregation, quantile aggregation and moment aggregation. As the dimension of the aggregated feature is still very high, we finally feed the aggregated feature into a linear regression model, known as partial least square regression (PLSR)~\cite{rosipal2006overview}, to predict the image quality.

Experiments are conducted on two realistic blur image databases (BID~\cite{ciancio2011no} and CLIVE~\cite{ghadiyaram2016massive}), as well as two synthetic blur image databases (TID2008~\cite{ponomarenko2009tid2008} and LIVE~\cite{sheikh2006statistical}). Our best proposal, named Semantic Feature Aggregation metric using PLSR (SFA-PLSR), is compared with the state-of-the-art methods. Experiments show that our method significantly outperforms the state-of-the-art on BID and CLIVE, and achieves comparable performance on TID2008 and LIVE. The good generalization ability of SFA-PLSR is validated by the cross dataset evaluation. We have also experimentally shown that high-level semantic features indeed play a more critical role than low-level features in resolving the challenging issue for NR-IQA of realistic blur images (see Figure~\ref{fig:Success}). This indicates a new perspective of blur perception in terms of the semantic loss.

The remainder of this paper is organized as follows. Section~\ref{sec:related} reviews the related work on NR-IQA of blur images. Section~\ref{sec:databases and criteria} introduces the benchmark databases and performance criteria. Section~\ref{sec:framework} describes our method in details. Section~\ref{sec:experiments} discusses the experimental results and analysis. And conclusions are made in Section~\ref{sec:conclusion}.

\section{Related Work}
\label{sec:related}
\subsection{Learning free methods}
Learning free methods use the characteristics of blur in terms of the spread of edges, the smoothing effects, the reduction of high frequency components or the loss of phase coherence.

The spread of edges can be used as a cue for blur estimation. Marziliano \textit{et al.}~\cite{marziliano2002no} used the average edge spread over all detected vertical Sobel edge locations as a quality metric for blur images. It can be further improved by incorporating the concept of just noticeable blur (JNB)~\cite{ferzli2009no} to adapt for the perception of human visual system (HVS). Since blur is not likely to be perceived when the edge width is small enough (below the width corresponding to JNB), Narvekar and Karam~\cite{narvekar2011no} assigned the quality score as the percentage of edges whose blur cannot be detected.

The smoothing effects of the blur process is useful information for NR-IQA. Gu \textit{et al.}~\cite{gu2015no} estimated image quality based on the energy-differences and contrast-differences of the locally estimated coefficients in the autoregressive parameter space. Bahrami and Kot~\cite{bahrami2014fast} considered the content based weighting distribution of the maximum local variation, which was modeled by the generalized Gaussian distribution (GGD). The estimated standard deviation was then used as an indicator of image quality. Later they also parameterized the image total variation distribution, and predicted image quality using the standard deviation modified by the shape-parameter to account for image content variation~\cite{bahrami2016catv}.

Image blur results in the reduction of high frequency components. Vu and Chandler~\cite{vu2012fast} estimated image quality using weighted average of the log-energies of the high-frequency coefficients. In~\cite{vu2012s3}, they generated a quality map based on a geometric mean of spectral and spatial measures. In view of the reduction of high-frequency components, the spectral measure was initially defined as the slope of the local magnitude spectrum, then rectified by a sigmoid function to account for HVS. To further consider the contrast effect, the spatial measure was calculated by the local total variation. Sang~\textit{et al.}~\cite{sang2014no} estimated image quality using the exponent of the truncated singular value curve of an image. Li~\textit{et al.}~\cite{li2016no} considered the moment energy, which can be affected by noticeable blur.

Blur also causes the loss of phase coherence, which gives a different perspective for understanding blur perception~\cite{wang2003local}. So Hassen~\textit{et al.}~\cite{hassen2013image} estimated image quality based on the strength of the local phase coherence near edges and lines.

\subsection{Learning-based methods}
Traditional learning free methods can not accurately express the diversity of blur process and the complexity of HVS. So recently machine learning technologies appear in IQA field. Learning-based methods mainly consist of two steps: feature extraction and quality prediction. In terms of feature extraction, these methods fall into two classes: the one using hand-crafted features and the other using learnt features.

Features can be manually designed using the natural scene statistics (NSS) of the image. NSS models of image coefficients in the spatial domain, wavelet domain and DCT domain are utilized in~\cite{mittal2012no,moorthy2011blind,saad2012blind} to extract quality relevant features, respectively. Tang~\textit{et al.}~\cite{tang2011learning} derived a set of low-level image quality features from NSS models, texture characteristics and blur/noise estimation. Ciancio~\textit{et al.}~\cite{ciancio2011no} used a neural network to combine eight existing methods and low-level features for blur image quality assessment. Oh~\textit{et al.}~\cite{oh2014no} evaluated image quality of camera-shaken images through mapping the spectral direction and shape features using support vector regression (SVR). Li~\textit{et al.}~\cite{li2016rise} took gradient similarity, singular value similarity and DCT domain entropies as quality features in a multi-scale framework. Li~\textit{et al.}~\cite{li2016blind} jointly considered the structural and luminance information in predicting image quality, where the structure information was described by the local binary pattern distribution and the normalized luminance magnitudes distribution portrayed the luminance information.

Machine learning techniques can learn quality relevant features. Li~\textit{et al.}~\cite{li2016image} and Lu~\textit{et al.}~\cite{Lu2016} extracted learnt features based on dictionary learning. Visual codebook is used to learn quality features in~\cite{ye2012unsupervised,xu2016blind}. Convolutional neural networks (CNN) have also been used to learn quality relevant features in NR-IQA~\cite{kang2014convolutional,bosse2016deep,yu2017cnn,kim2017fully,siahaan2016augmenting,sun2016no}. Kang~\textit{et al.}~\cite{kang2014convolutional} integrated feature learning and patch quality prediction into an end-to-end network, and the image quality was estimated by the average score of all sampling patches. Following~\cite{kang2014convolutional}, the network was deeper and weights for patch scores were also integrated into the learning process~\cite{bosse2016deep}. In~\cite{yu2017cnn}, CNN was used to learn features and the general regression neural network was used as the predictor. In~\cite{kim2017fully}, a sub-network was first trained on patches using the FR-IQA scores, and then a whole network from images to quality was trained.

The most related works to ours are~\cite{siahaan2016augmenting,sun2016no}, which resize the image to meet the required input size of the pre-trained AlexNet so as to extract the image features. Our work differs from  them mainly in three ways: (1) Unlike~\cite{siahaan2016augmenting,sun2016no}, we use multiple overlapping image patches instead of the resized image to represent the image, which can avoid introducing deformation as well as cover the image information. Correspondingly, we propose three effective statistical structures to conduct feature aggregation. (2) The features extracted from the pre-trained DCNN model  in~\cite{siahaan2016augmenting,sun2016no} are only used as the auxiliary to boost the performance of methods based on low-level features, while our aggregated semantic features are directly used as quality relevant features. (3) We focus on realistic blur, and since residual images contain important cues about image blur, the residual learning based network (ResNet-50~\cite{he2016deep}) is selected as the feature extractor instead of the one in~\cite{siahaan2016augmenting,sun2016no} without residual learning.

\section{Benchmark Databases and Performance Criteria}
\label{sec:databases and criteria}
\subsection{Benchmark Databases}
In this work, we consider two realistic image databases (BID~\cite{ciancio2011no} and CLIVE~\cite{ghadiyaram2016massive}), as well as two synthetic blur image datasets from TID2008~\cite{ponomarenko2009tid2008} and LIVE~\cite{sheikh2006statistical}.

\textbf{BID} includes totally 586 realistic blur images taken from real world along with a variety of scenes, light conditions, camera apertures and exposure time. Subjective quality scores are provided in the form of mean opinion score (MOS) ranging from 0 to 5.

\textbf{CLIVE} includes 1162 realistic distorted images captured using real-world mobile cameras, most of which suffer from motion blur or out-of-focus blur. Subjective quality scores are provided in the form of MOS ranging from 0 to 100.

\textbf{TID2008} contains 1700 distorted images, in which we only consider the 100 Gaussian blur images. There are only 25 reference images, and 4 blur kernels for each reference image. Subjective quality scores are provided in the form of MOS ranging from 0 to 9.

\textbf{LIVE} contains 779 distorted images, in which we only consider the 145 Gaussian blur images. There are only 29 reference images, and 5 blur kernels for each reference image. Subjective quality scores are provided in the form of Difference of MOS (DMOS) ranging from 0 to 100.

\subsection{Performance Evaluation Criteria}
Three evaluation criteria are chosen to evaluate the performance of NR-IQA methods: Spearman's rank-order correlation coefficient (SROCC), Pearson's linear correlation coefficient (PLCC) and root mean square error (RMSE). PLCC and RMSE are used for measuring prediction accuracy, while SROCC is used for measuring prediction monotonicity. For these three criteria, larger PLCC/SROCC and smaller RMSE indicate better performance. Before calculating PLCC and RMSE values of the learning free methods, a nonlinear fitting is needed to map the objective scores to the same scales of the subjective scores. In this paper, we adopt the following four-parameter logistic function recommended in~\cite{vqeg2000fr}.

\begin{equation}
f(x)=\frac{\tau_1-\tau_2}{1+e^{\frac{x-\tau_3}{\tau_4}}}+\tau_2
\end{equation}
where $\tau_1$ to $\tau_4$ are free parameters to be determined during the curve fitting process.

Monte-Carlo cross validation is used for learning-based methods. For each database, 80\% data are for training and 20\% data are for testing. There is no same ``original images" between training data and testing data. This procedure is repeated 1000 times and the median or mean values are reported. It should be noted that the learning free methods are tested on the same data as learning-based methods. \ldq{Besides, we should specifically point out that the training data on BID are used in Section~\ref{sec:framework} for the comparative study.}

\section{The Proposed Method}
\label{sec:framework}
\begin{figure*}[!htb]
	\centering
	\includegraphics[width=0.85\textwidth]{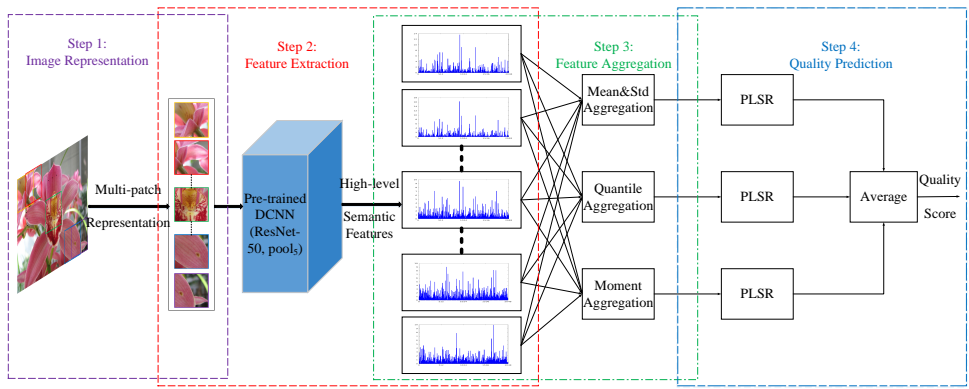}
	\caption{The overall framework of the proposed method, mainly includes four steps: image representation, feature extraction, feature aggregation, and quality prediction.}
	\label{fig:framework}
\end{figure*}

\ldq{The framework of the proposed method is shown in Figure~\ref{fig:framework}, including four steps: image representation, feature extraction, feature aggregation, and quality prediction. In this section, we will conduct an in-depth comparative study to determine the best choice for each step.}

\subsection{Image Representation}
The pre-trained DCNN models (\textit{e.g.}, AlexNet) require a fixed input size. To meet this requirement, images can be cropped, or resized to the fixed size. Since the resizing operation can introduce geometric deformation, which may change the image quality, it is not a good way. In the mean time, cropping only the central patch is not enough to cover the information of a large image. Because of these two issues, we consider using multiple overlapping patches to represent the image, which not only covers information of the whole image but also avoids introducing geometric deformation.

We compare the impact of four different image representations, including the cropping, scaling, padding and multi-patch representation. Cropping representation uses the central patch to represent the image. Padding representation preserves the aspect ratio by resizing the larger dimension to the required length and then padding zeros to the smaller dimension. Scaling representation directly resizes the image without keeping the original aspect ratio. Multi-patch representation generates multiple overlapping patches that are uniformly sampled over the whole image with a sampling stride\footnote{There is no significant performance variation among different sampling strides when it is subjected to cover the whole information, so the sampling stride is simply fixed to be half of the patch size.}.

\begin{figure}[!htb]
	\centering
	\includegraphics[width=2in]{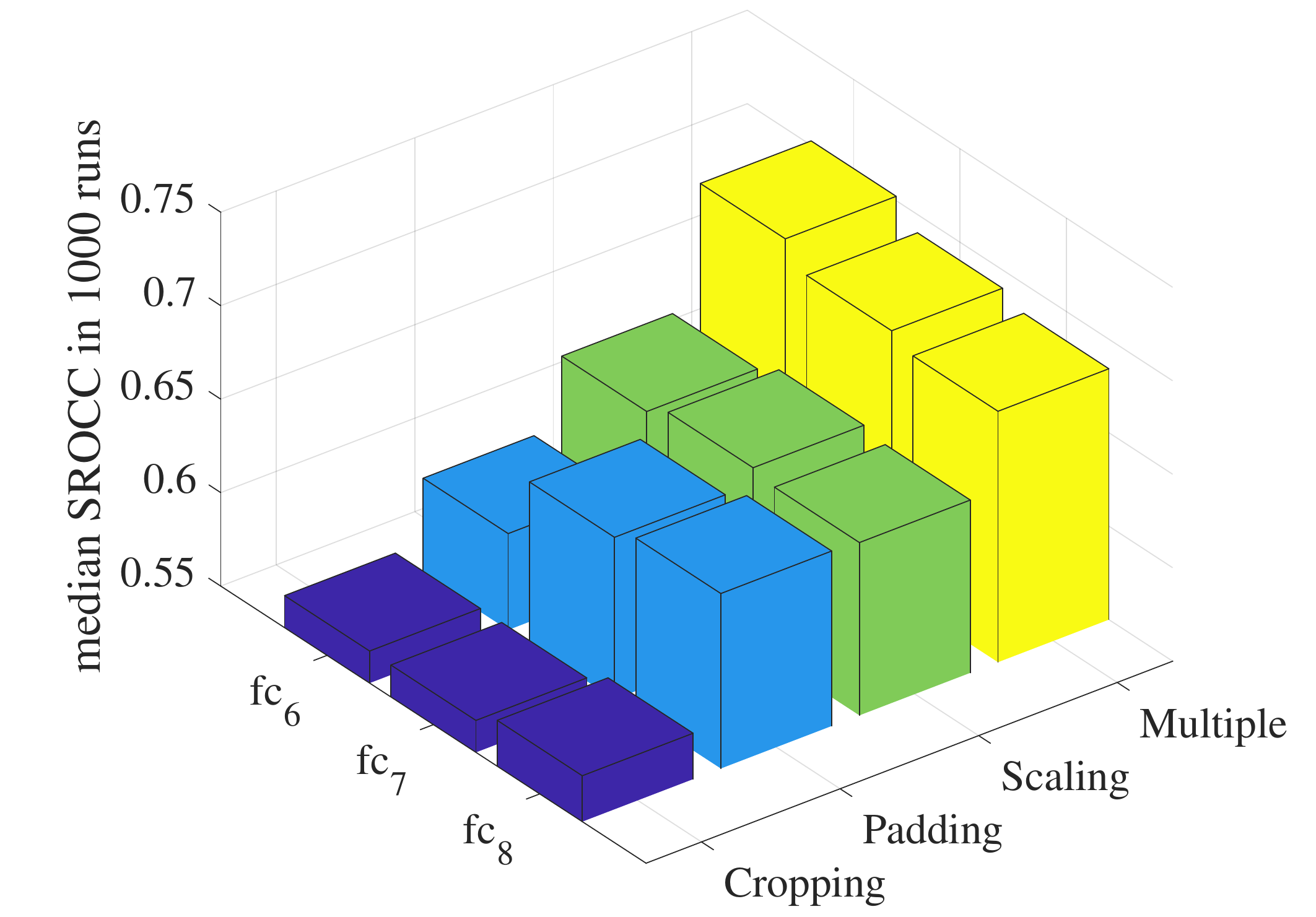}
	\caption{Comparison of different image representations. No matter which layer is used to extract features, the multi-patch representation achieves the best performance.}
	\label{fig:ImageRepresentation}
\end{figure}

\ldq{To perform the comparison, we need a baseline for feature extraction, feature aggregation and quality prediction. Before the comparative study on the following steps, we choose the classical pre-trained DCNN model AlexNet and extract features from the frequently-used fully connected layers (\textit{i.e.}, \(fc_6\), \(fc_7\) and \(fc_8\)). For feature aggregation, we choose the mean feature vector for simplicity. PLSR is used for quality prediction. The comparative study is conducted on the training data of BID, where 20\% of the training data are used as validation data and the performance on the validation set is used for comparison (the same below).} It can be seen from Figure~\ref{fig:ImageRepresentation} that (1) cropping representation obtains the worst performance, (2) since resizing operation keeps most of the image information, padding and scaling representation achieve better performance than cropping representation, (3) the use of multi-patch representation significantly outperforms the other three. So we decide to use the multi-patch representation in our framework.

\subsection{Feature Extraction}
Given an image \(\mathbf{I}\), we represent it with a set of multiple overlapping patches \(\{\mathbf{p}_1,\cdots,\mathbf{p}_n\}\), and then feed these patches into an off-the-shelf DCNN model to extract features. For each patch \(\mathbf{p}_j\), the extracted feature is denoted by
\begin{equation}
\mathbf{d}_j=DCNN(\mathbf{p}_j,L;\theta), \, j=1,\cdots,n.
\end{equation}
where \(L\) indicates which layer (\textit{e.g.}, ${fc}_8$ layer in AlexNet) to extract features and  \(\theta\) is the trained network parameter.

\begin{figure}[!htb]
	\centering
	\includegraphics[width=3.2in]{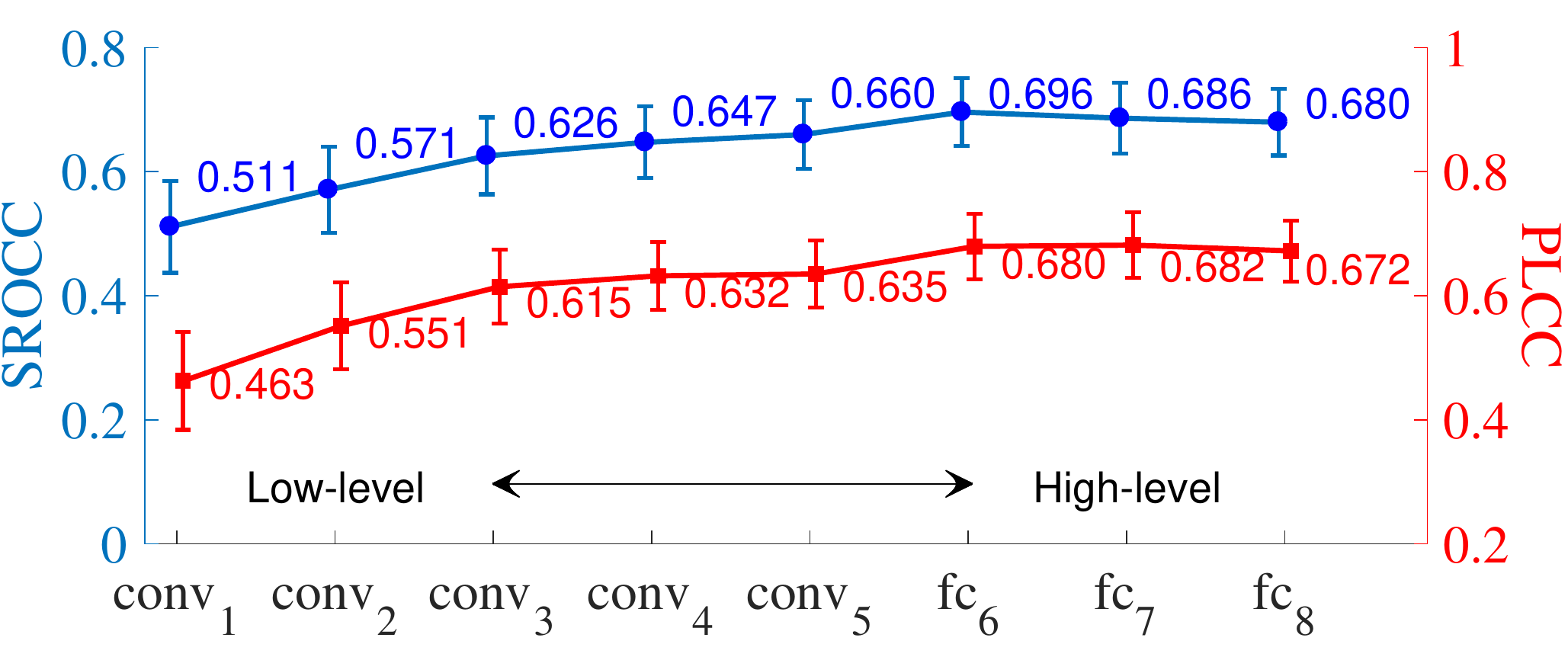}
	\caption{[Best viewed in color.] Mean and standard deviation of SROCC and PLCC. x-axis indicates from which layer (in AlexNet) we extract features. The curve indicates the mean values and the error bars indicate the standard deviations.}
	\label{fig:Feature}
\end{figure}

\textbf{The role of high-level semantics}: Pre-trained DCNN models for image classification or scene recognition have encoded semantics in high-level features. Here, we conduct a comparative study to investigate the role of high-level semantics in NR-IQA of realistic blur images. We take the AlexNet as the pre-trained model, and extract features of multiple patches from its different layers\footnote{Since the response of the convolutional layer is a set of feature maps, we derive features by global average pooling.} (\({conv}_1\) to \({conv}_5\) and \({fc}_6\) to \({fc}_8\)). PLSR maps the mean feature vector to the quality score. From the plot in Figure~\ref{fig:Feature}, we have the following observations. First, high-level features are better than the low-level features, which indicates that high-level semantic features play an important role in NR-IQA of realistic blur images. However, the feature extracted from the top layer (${fc}_8$) is slightly worse than the second and third top layers (${fc}_7,{fc}_6$). This is because the top layer is directly linked to the classifier and the extracted feature is task-specific, which may contain only the classification information. The third top layer (${fc}_6$), close to the last convolutional layer, achieves the best performance in terms of SROCC. Therefore, in our framework, we consider the second or third top layer close to the last convolutional layer to extract features.

\begin{figure}[!htb]
	\centering
	\includegraphics[width=2in]{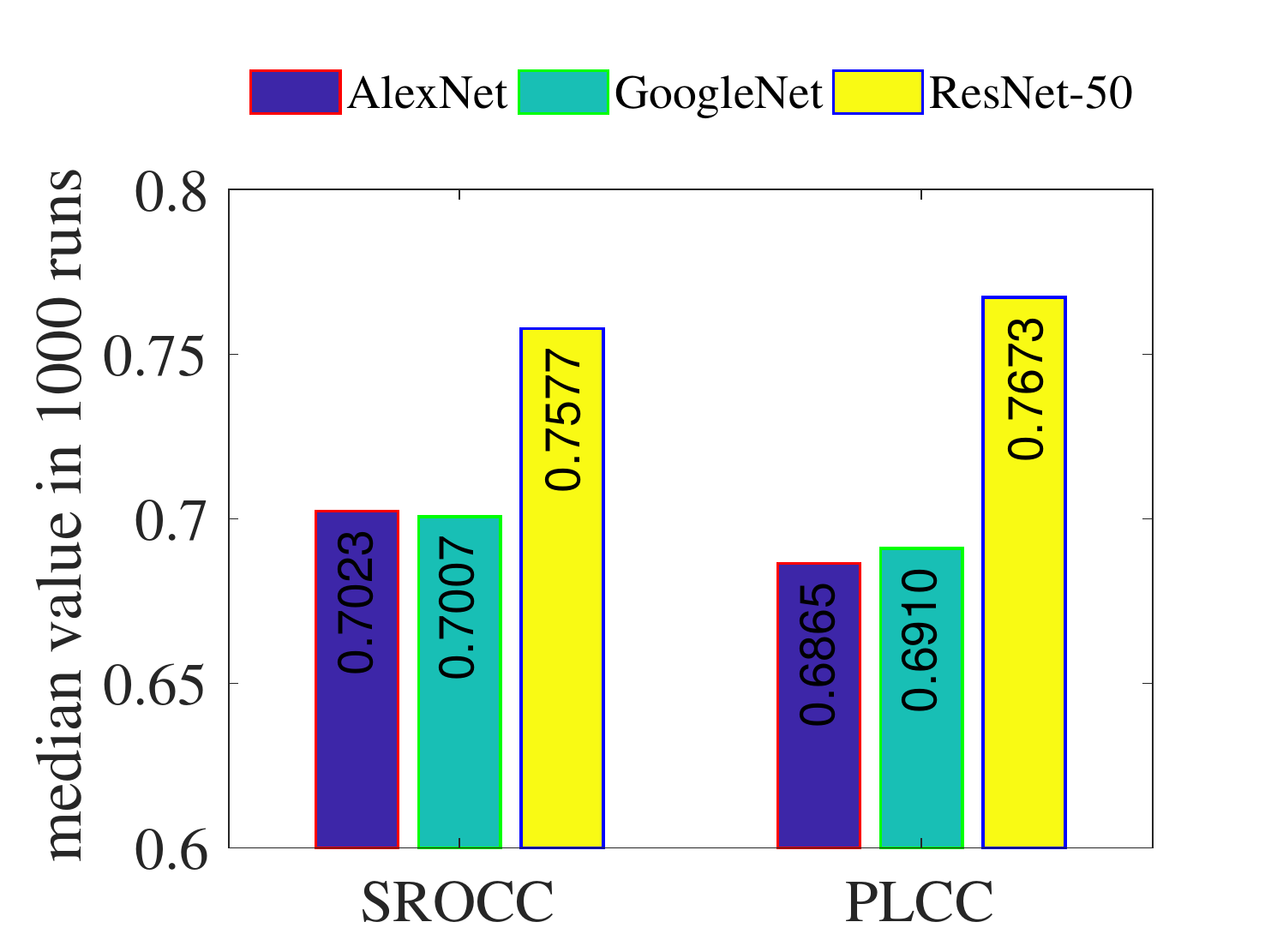}
	\caption{Comparison of different pre-trained DCNN models.}
	\label{fig:PretrainedModel}
\end{figure}

\textbf{Impact of different pre-trained DCNN models}: We also compare different pre-trained DCNN models in the proposed framework, including AlexNet~\cite{krizhevsky2012imagenet}, GoogleNet~\cite{szegedy2015going} and ResNet-50~\cite{he2016deep}, where the features are extracted from the \({fc}_6\), \({pool}_5/7\times7\_s_1\) and \({pool}_5\) layer, respectively. The quality prediction step is still based on PLSR. Figure~\ref{fig:PretrainedModel} shows the performance values, from which we can observe that ResNet-50 achieves the best performance. It is shown that the residual image contains important information in capturing quality relevant features~\cite{yan2016blind}. Besides, image blur can be more easily captured in residual images. So the significant gain in ResNet-50 may due to the residual learning, and we choose ResNet-50 as the feature extractor.

\subsection{Feature Aggregation}
With the extracted features, we need to aggregate them into a single one. One straightforward way is to concatenate all these \(n\) features into a long feature vector, \textit{i.e.},

\begin{equation}
\mathbf{f}_{concat}=\mathbf{d}_1\oplus\cdots\mathbf{d}_j\cdots\oplus\mathbf{d}_n
\end{equation}
where \(\oplus\) is the concatenation operator.

However, it will result in a very high dimension of the feature space. Besides, the dimension of the concatenated feature vector will depend on the number of patches, which is not the same among the images with different resolutions. To avoid this, we can take the mean value of all features in each dimension, that is,

\begin{equation}
\mathbf{f}_{mean}=\left(m_1,\cdots,m_i,\cdots,m_l\right)^T,
\end{equation}
\[m_i=\frac{\sum_{j=1}^nd_{ji}}{n},i=1,\cdots,l.\]
where \(d_{ji}\) is the i-th element of \(\mathbf{d}_j\) and \(l\) is the dimension of  \(\mathbf{d}_j\).

The mean aggregation structure loses important information (\textit{e.g.}, the standard deviation in each dimension) of the feature set. So we propose three different statistical structures for feature aggregation, namely, mean\&std aggregation, quantile aggregation and moment aggregation.

\textbf{Mean\&std aggregation}: The standard deviation in each dimension is further considered, and the first aggregated feature \(\mathbf{f}_1\) is obtained by:
\begin{equation}
\mathbf{f}_1 = \mathbf{f}_{mean}\oplus\mathbf{f}_{std}
\end{equation}
\[\mathbf{f}_{std} = \left(\sqrt{\frac{\sum_{j=1}^n(d_{j1}-m_{1})^2}{n-1}},\cdots,\sqrt{\frac{\sum_{j=1}^n(d_{jl}-m_{l})^2}{n-1}}\right)^T\]
where \(m_i\) is the \(i\)-th element of \(\mathbf{f}_{mean},i=1,\cdots,l.\)

\textbf{Quantile aggregation}: Quantiles are important order statistics. We consider the widely used quartiles. The min, the median and the max are the zeroth, second, and fourth quartile, respectively. We denote the zeroth to fourth quartile of \((d_{1i},\cdots,d_{ni})\) as \(d^{(0)}_{i},d^{(1)}_{i},d^{(2)}_{i},d^{(3)}_{i},d^{(4)}_{i},i=1,\cdots,l\), respectively. So the second aggregated feature \(\mathbf{f}_2\) based on quartiles can be defined as:
\begin{equation}
\mathbf{f}_2 = \mathbf{q}_{0}\oplus\mathbf{q}_{1}\oplus\mathbf{q}_{2}\oplus\mathbf{q}_{3}\oplus\mathbf{q}_{4}
\end{equation}
\[\mathbf{q}_{t} = \left(d^{(t)}_{1},\cdots,d^{(t)}_{l}\right)^T,t=0,1,2,3,4.\]

\textbf{Moment aggregation}: Moments also play an important role in describing the statistics of a distribution. Mean is actually the origin moment of first-order. In order to balance between the need of more information and the dimension reduction of the feature space, we further consider the \(k\)-th root of the central moment of order \(k\,(k=2,3,4)\)\footnote{Note that the first central moment is zero, and here the second central moment is the variance computed using a divisor of \(n\) rather than \(n-1\).}, and obtain the third aggregated feature \(\mathbf{f}_3\):
\begin{equation}
\mathbf{f}_3 = \mathbf{f}_{mean}\oplus\mathbf{M}_{2}\oplus\mathbf{M}_{3}\oplus\mathbf{M}_{4}
\end{equation}
\[\mathbf{M}_{k} = \left(\sqrt[k]{\frac{\sum_{j=1}^n(d_{j1}-m_{1})^k}{n}},\cdots,\sqrt[k]{\frac{\sum_{j=1}^n(d_{jl}-m_{l})^k}{n}}\right)^T\]
where \(k=2,3,4.\) \(m_i\) is the \(i\)-th element of \(\mathbf{f}_{mean},i=1,\cdots,l.\)

\begin{figure*}[!htb]
	\centering
	\includegraphics[width=0.8\textwidth]{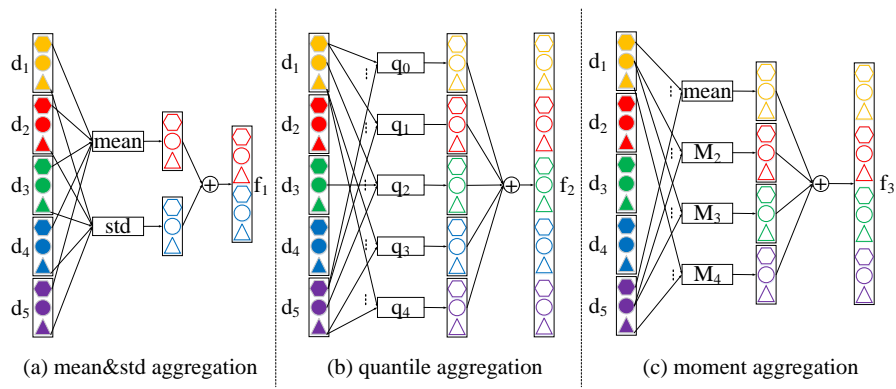}
	\caption{[Best viewed in color.] An example of the three statistical structures for feature aggregation. The input is $n = 5$ features $\{\mathbf{d}_1, \mathbf{d}_2, \mathbf{d}_3,\mathbf{d}_4,\mathbf{d}_5\}$, where the feature dimension is $l = 3$. \((\mathbf{q}_{0},\mathbf{q}_{1},\mathbf{q}_{2},\mathbf{q}_{3},\mathbf{q}_{4})\) indicate the five quartiles, and \(\mathbf{M}_{k}\) represents the \(k\)-th root of the central moment of order \(k\) (\(k=2,3,4\)). Not all the connections are shown between input and statistical functions for clarity.}
	\label{fig:aggregation}
\end{figure*}

The aforementioned three statistical structures for feature aggregation result in a $2l, 5l$ and $4l$-dimensional feature vector, respectively. An example of these aggregation structures when $n=5, l=3$ is shown in Figure~\ref{fig:aggregation}.

\subsubsection{Contribution of Different Statistical Aggregation Structures}
We compare the mean aggregation (baseline) with the three proposed statistical aggregation structures. The ResNet-50 is used as the feature extractor of multiple patches and PLSR is used as the regression model. Table~\ref{tab:Aggregation} summarizes the median values of SROCC, PLCC and RMSE. The best result comes from the ensemble of the three statistical structures and has been marked in boldface. We can see that the three proposed statistical structures have significant gain over the baseline, from which we verify the effectiveness of the proposed aggregation structures on capturing the information of the feature set.

\begin{table}[!htb]
	\centering
	\caption{Comparison among different aggregation structures. ``average-quality" means averaging scores of the three proposed structures.}
	\label{tab:Aggregation}
	\begin{small}
	\begin{threeparttable}
	\begin{tabular}{lcccc}
	\toprule
	Aggregated Feature & SROCC & PLCC & RMSE \\
	\midrule
	mean (\(\mathbf{f}_{mean}\)) & 0.7577 & 0.7673 & 0.8283 \\
	\midrule
	mean\&std (\(\mathbf{f}_{1}\)) & 0.8022 & 0.8174 & 0.7333\\
    quantile (\(\mathbf{f}_{2}\)) & 0.8109 & 0.8254 & 0.7135\\
	moment (\(\mathbf{f}_{3}\)) & 0.8100 & 0.8254 & 0.7171\\
    \(\mathbf{f}_{1}\oplus\mathbf{f}_{2}\) & 0.8123 & 0.8269 & 0.7116\\
    \(\mathbf{f}_{3}\oplus\mathbf{f}_{2}\) & 0.8127 & 0.8270 & 0.7121\\
    average-quality (\(\mathbf{f}_{1},\mathbf{f}_{2},\mathbf{f}_{3}\)) & \textbf{0.8154} & \textbf{0.8305} & \textbf{0.7055}\\
	\bottomrule
	\end{tabular}
	\end{threeparttable}
	\end{small}
\end{table}

\subsection{Quality Prediction}
With the help of statistical structures for feature aggregation, we reduce the dimension of feature space (\(nl\rightarrow 2l, 4l, 5l\)) and make the dimension independent of the number of patches. However, in the pre-trained DCNN, \(l\) is also a large number (\(l=2048\) in ResNet-50's ${pool}_5$ layer). Since the dimension of the feature space is much larger than the number of our training samples, we consider the linear regression model. Specifically, partial least square regression (PLSR)~\cite{rosipal2006overview} is adopted in our work because of its low-complexity and remarkable capability to handle high-dimensional data. PLSR  reduces the input high-dimensional features to several uncorrelated latent components and then performs least squares regression on these components. There is only one parameter \(p\) (the number of components) in PLSR, which can be determined by cross validation.


After the above investigations, we obtain our best proposal, dubbed as Semantic Feature Aggregation metric using PLSR (SFA-PLSR). It uses multiple overlapping patches to represent images, and extracts features from the \({pool}_5\) layer of the pre-trained ResNet-50 model, as well as averages the scores of the mean\&std aggregation, quantile aggregation and moment aggregation.
\section{Experiments}
\label{sec:experiments}

In the following parts, we compare the performance of the proposed SFA-PLSR method with the state-of-the-art NR-IQA methods in both intra-database and inter-database scenarios. As for the software platform to implement our proposed method, we use the Caffe~\cite{jia2014caffe} framework to extract the features from the pre-trained DCNN model. PLSR is performed by the MATLAB function \textit{plsregress}, and its parameter \(p\) is globally set to \(10\) based on the 5-fold cross-validation using the training data of a single run (on BID), where p is selected from the set $\{5, 10, 15, 20, 25, 30\}$ for simplicity.

\subsection{Performance Comparison}
In this part, we compare the performance of SFA-PLSR with ten existing (from classical to the most up to date) NR-IQA methods of blur images, which are MDWE~\cite{marziliano2002no}, CPBD~\cite{narvekar2011no},FISH~\cite{vu2012fast}, S3~\cite{vu2012s3}, LPC~\cite{hassen2013image}, MLV~\cite{bahrami2014fast}, ARISM~\cite{gu2015no}, BIBLE~\cite{li2016no}, SPARISH~\cite{li2016image} and RISE~\cite{li2016rise}. Five remarkable general-purpose NR-IQA methods, including BRISQUE~\cite{mittal2012no}, Kang's CNN~\cite{kang2014convolutional}, FRIQUEE~\cite{ghadiyaram2016perceptual}, NRSL~\cite{li2016blind}, and S-HOSA~\cite{siahaan2016augmenting}, are also taken for comparison.

Table~\ref{tab:performance} reports the median SROCC, PLCC and RMSE in 1000 runs on the four databases. \ldq{We also report the weighted-average SROCC over all four databases as the overall performance, where the weights are proportional to the database-sizes (see the last column of Table~\ref{tab:performance}).} Among the ten NR-IQA methods of blur images, the first nine methods fail on the two realistic databases (SROCC\(<0.5\) on BID and CLIVE) due to their neglect of global semantic information, and RISE achieves the best performance on BID. The proposed method SFA-PLSR significantly outperforms others on BID and CLIVE in both prediction accuracy (PLCC, RMSE) and monotonicity (SROCC). As for the general purpose NR-IQA methods, FRIQUEE and S-HOSA achieve better performance on the realistic databases than the others. Kang's CNN~\cite{kang2014convolutional} does not perform well because it assumes that patch quality equals to image quality, which is not true for these two realistic image datasets. On TID2008 and LIVE, there are less than 30 images with different contents, which is much smaller than BID (586) and CLIVE (1162), so the role of semantic information is weakened and the impact of low-level features is enhanced. Nevertheless, our method SFA-PLSR still achieves comparable performance on the two synthetic databases. \ldq{In general, our method also achieves the best overall performance}.

\begin{table*}[!hbt]
	\centering
	\caption{Performance comparison on four databases. In each column, the best performance value is marked in boldface and the second best performance value is underlined. The last column indicates the weighted-average of SROCC over all four databases, where the weights are proportional to the database-sizes.}
	\label{tab:performance}
	\begin{small}
	\resizebox{\columnwidth}{!}{
	\begin{threeparttable}
	\begin{tabular}{ll|ccc|ccc|c|c|c}
	\toprule
	\multirow{2}{*}{Category} & \multirow{2}{*}{Method} & \multicolumn{3}{c|}{BID~\cite{ciancio2011no}} & \multicolumn{3}{c|}{CLIVE~\cite{ghadiyaram2016massive}} & TID2008~\cite{ponomarenko2009tid2008} & LIVE~\cite{sheikh2006statistical} & Overall\\
	& & SROCC & PLCC & RMSE & SROCC & PLCC & RMSE & SROCC & SROCC & SROCC \\
	\midrule
	\multirow{10}{1.2cm}{NR-IQA of blur images} & MDWE~\cite{marziliano2002no} & 0.3067 & 0.3538 & 1.1639 & 0.4313 & 0.4988 & 17.5025 & 0.8556 & 0.9188 & 0.4514\\
	& CPBD~\cite{narvekar2011no} & 0.0202 & 0.2181 & 1.2166 & 0.3027 & 0.4026 & 18.4602 & 0.8723 & 0.9390 & 0.2945\\
	& FISH~\cite{vu2012fast} & 0.4736 & 0.4853 & 1.0894 & 0.4865 & 0.5380 & 17.0310 & 0.8737 & 0.9008 & 0.5323\\
	& S3~\cite{vu2012s3} & 0.4109 & 0.4471 & 1.1177  & 0.4034 & 0.4864 & 17.6224 & 0.8650 & 0.9515 & 0.4686\\
	& LPC~\cite{hassen2013image} & 0.3150 & 0.4053 & 1.1408 & 0.1483 & 0.3490 & 18.9205 & 0.8805 & 0.9469 & 0.2922\\
	& MLV~\cite{bahrami2014fast} & 0.3169 & 0.3750 & 1.1561 & 0.3412 & 0.4076 & 18.4350 & 0.8977 & 0.9431 & 0.4058\\
	& ARISM~\cite{gu2015no} & 0.0151 & 0.1929 & 1.2245 & 0.2427 & 0.3554 & 18.8947 & 0.8851 & 0.9585 & 0.2601\\
	& BIBLE~\cite{li2016no} & 0.3609 & 0.3923 & 1.1469 & 0.4260 & 0.5178 & 17.3007 & 0.9114 & \textbf{0.9638} & 0.4703\\
	& SPARISH~\cite{li2016image} & 0.3071 & 0.3555 & 1.1659 & 0.4015 & 0.4843 & 17.6702 & 0.9126 & \textbf{0.9638} & 0.4403\\
	& RISE~\cite{li2016rise} & 0.5839 & 0.6017 & 0.9936 & - & - & - & \underline{0.9218} & 0.9493 & 0.6833\\
	\midrule
	Proposed & \textbf{SFA-PLSR} & \textbf{0.8269} & \textbf{0.8401} & \textbf{0.6854} & \textbf{0.8130} & \textbf{0.8313} & \textbf{11.3905} & 0.9098 & 0.9523 & \textbf{0.8321}\\
	\midrule
	\multirow{5}{1.2cm}{General purpose NR-IQA} & BRISQUE~\cite{mittal2012no} & 0.5795 & 0.5754 & 1.0624 & 0.5950 & 0.6195 & 16.0273 & 0.8737 & 0.8892 & 0.6258\\
	& Kang's CNN~\cite{kang2014convolutional} & 0.4818 & 0.4977 & 1.1030 & 0.4964 & 0.5218 & 17.8567 & 0.9000 & 0.9429 & 0.5448\\
	& FRIQUEE~\cite{ghadiyaram2016perceptual} & \underline{0.7359} & \underline{0.7477} & \underline{0.8433} & 0.6916 & 0.7069 & 14.4244 & \textbf{0.9261} & 0.9515 & \underline{0.7353}\\
	& NRSL~\cite{li2016blind} & 0.638 & 0.663 & 0.931 & 0.631 & 0.654 & 15.317 & - & \underline{0.959} & 0.658\\
	& S-HOSA~\cite{siahaan2016augmenting} & 0.6869 & 0.6913 & 0.9112 & \underline{0.7051} & \underline{0.7241} & \underline{14.0237} & 0.8729 & 0.9469 & 0.7258\\
	\bottomrule
	\end{tabular}
	\begin{tablenotes}
        \footnotesize
        \item[] The results of RISE and NRSL are from their \textbf{original papers}. The code of Kang's CNN and S-HOSA are written by ourselves following the detail of their papers, and the codes of other compared methods are from \textbf{original authors}.
    \end{tablenotes}
    \end{threeparttable}
    }
	\end{small}
\end{table*}

\subsection{Cross Dataset Evaluation}
In this subsection, we test the generalization capability of learning-based methods through cross dataset evaluation. Since learning-based methods assume testing images and training images have a similar distribution, we conduct cross dataset evaluation on realistic databases (BID and CLIVE) and synthetic databases (TID2008 and LIVE), respectively. It should be noted that CLIVE contains 383 images resized from BID images, we exclude the 383 images from CLIVE in cross dataset experiments.

We compare our method with RISE (the compared NR-IQA method of blur images with the best overall performance), FRIQUEE and S-HOSA (the best two general purpose NR-IQA methods). The SROCC values are provided in Table~\ref{tab:cross dataset evaluation}. It can be seen that our method performs better than RISE, FRIQUEE and S-HOSA, which has demonstrated the database independency and robustness of the proposed SFA-PLSR method.

\begin{table}[!htb]
	\centering
	\caption{SROCC values in cross dataset evaluation.}
	\label{tab:cross dataset evaluation}
	\begin{small}
	\begin{threeparttable}
	\begin{tabular}{cccccc}
	\toprule
	train \(\rightarrow\) test & RISE & FRIQUEE & S-HOSA & Ours \\
	\midrule
	BID \(\rightarrow\) CLIVE & - & 0.3571 & 0.4767 & \textbf{0.5729} \\
	CLIVE \(\rightarrow\) BID & - & 0.3886 & 0.3433 & \textbf{0.6838}\\
    TID2008 \(\rightarrow\) LIVE & 0.8638 & 0.8690 & 0.8950 & \textbf{0.9166}\\
	LIVE \(\rightarrow\) TID2008 & 0.9138 & 0.8727 & 0.8612 & \textbf{0.9243}\\
	\bottomrule
	\end{tabular}
	\begin{tablenotes}
        \footnotesize
        \item[] The results of cross dataset evaluation on the two realistic blur datasets were not reported in the original paper of RISE.
    \end{tablenotes}
    \end{threeparttable}
	\end{small}
\end{table}

\subsection{Impact of Training Ratio}
\begin{figure}[!htb]
	\centering
	\includegraphics[width=2in]{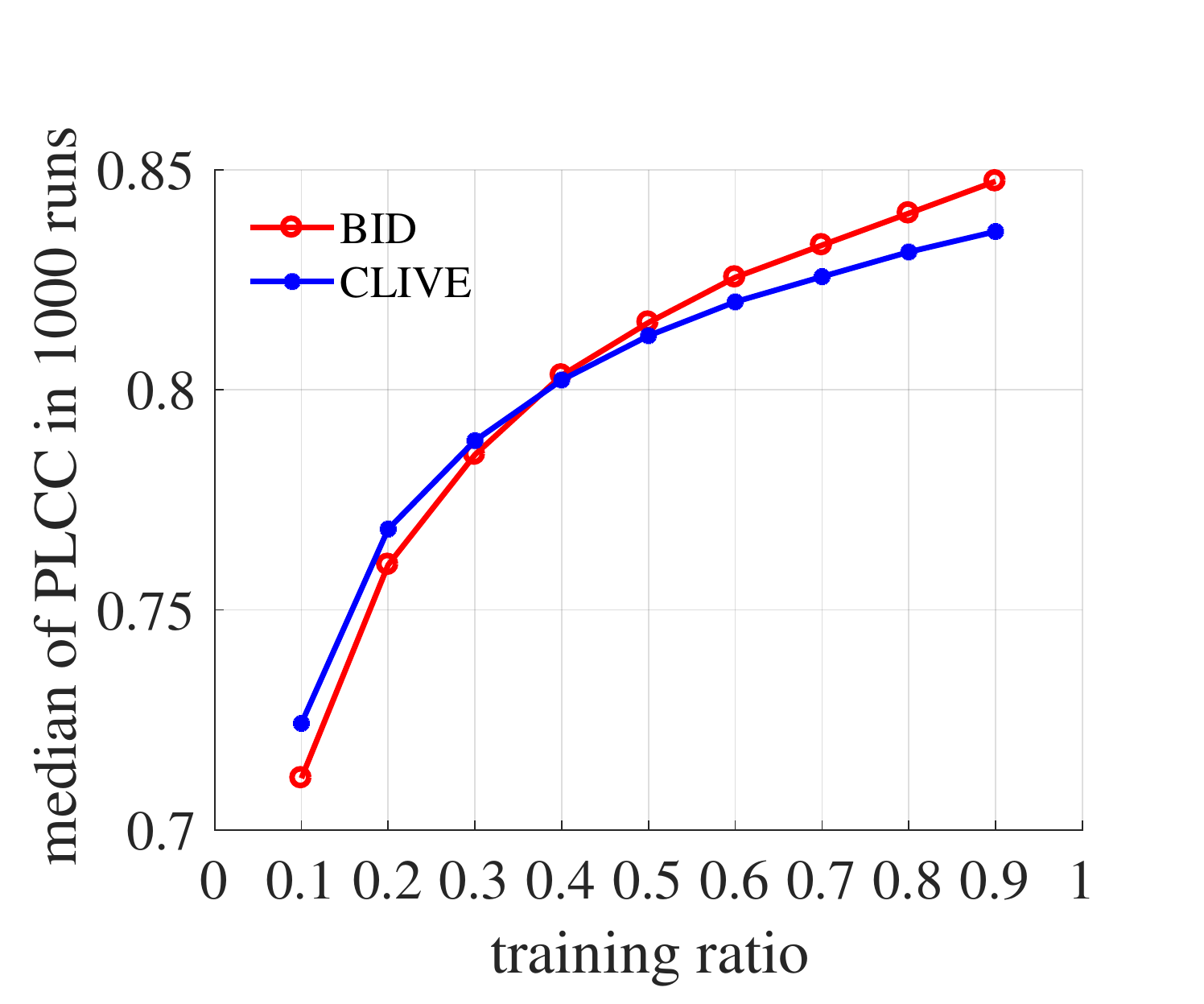}
	\caption{The PLCC of SFA-PLSR with different training ratios.}
	\label{fig:TrainingRatio}
\end{figure}

In order to have an intuitive understanding of how the training ratio affects the performance of our methods, we also conduct an experiment to test SFA-PLSR with different training ratios (from 10\% to 90\% with an increment step of 10\%). It is clearly shown from Figure~\ref{fig:TrainingRatio} that with the increase of training ratio, the performance values boost quickly when the training ratio is smaller than 30\%. We can see that even if only 40\% of images are used for training, the PLCC values are still close to 0.8. This is helpful in real-world applications, where relatively small amount of images are labeled.

\subsection{$2\sigma$-Confidence Band and Failure Case}
\begin{figure}[!htb]	
	\centering
	\begin{subfigure}{1.8in}
		\centering
		\includegraphics[width=1.8in]{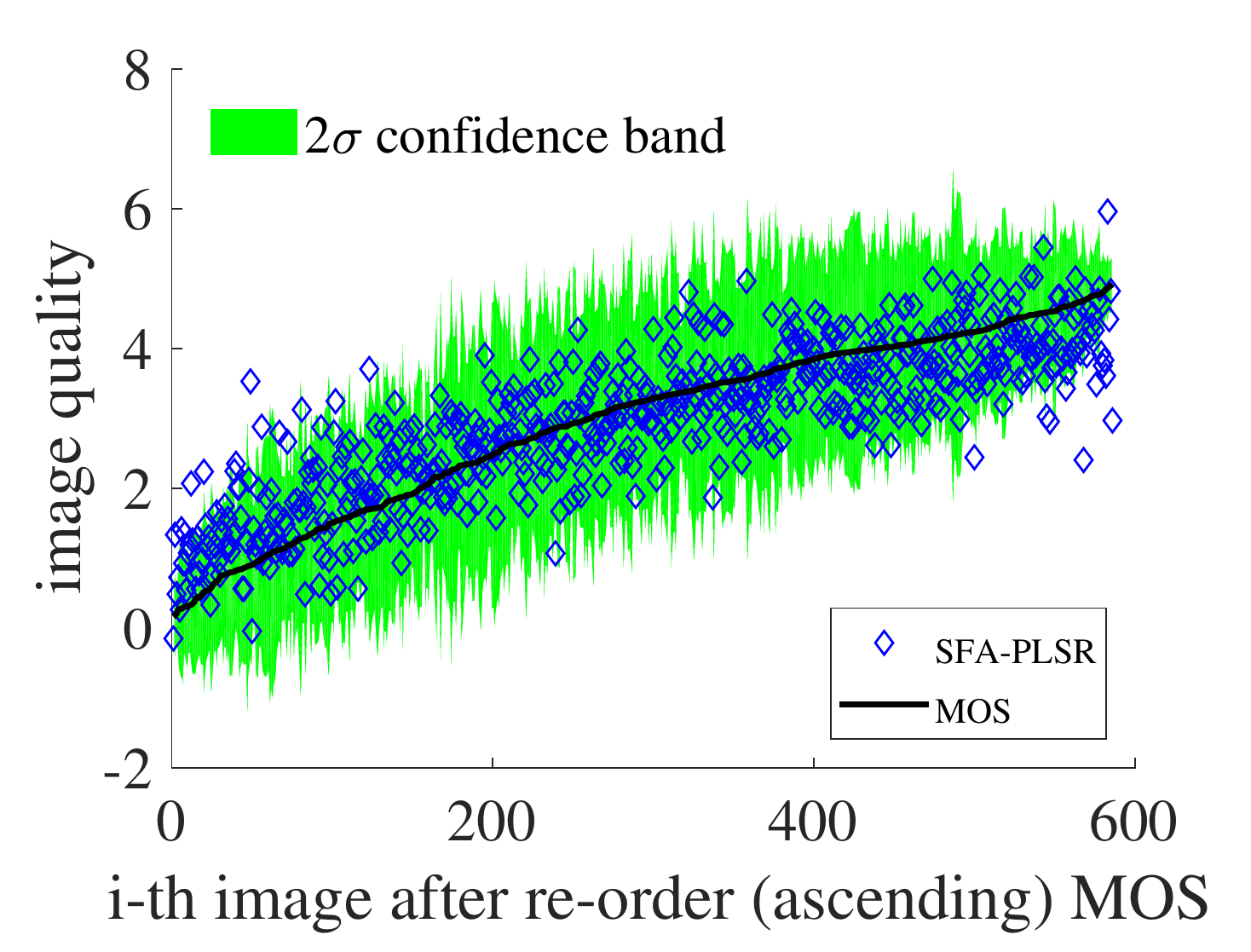}
		\caption{SFA-PLSR}
	\end{subfigure}
	\begin{subfigure}{1.8in}
		\centering
		\includegraphics[width=1.8in]{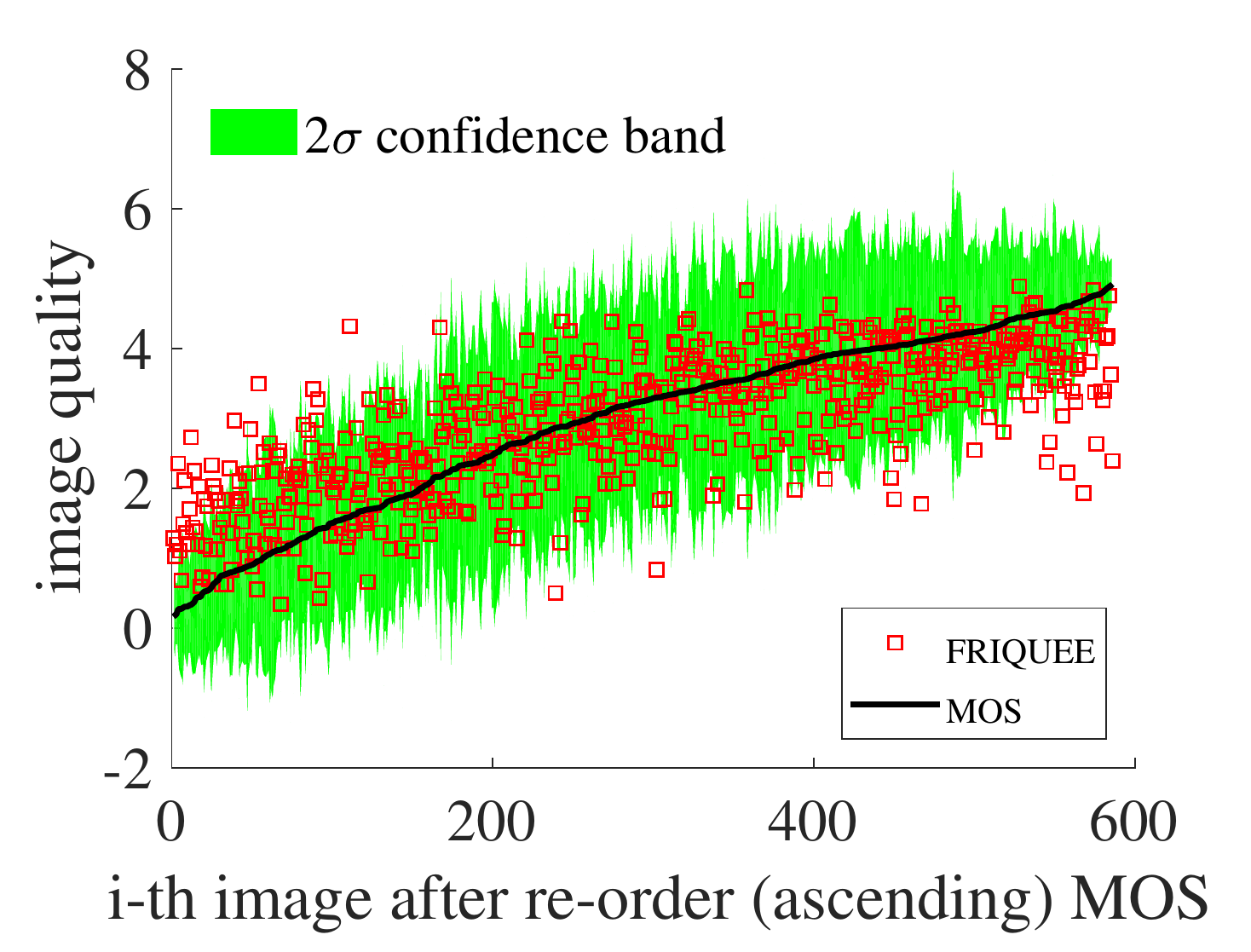}
		\caption{FRIQUEE}
	\end{subfigure}
	\begin{subfigure}{1.8in}
		\centering
		\includegraphics[width=1.8in]{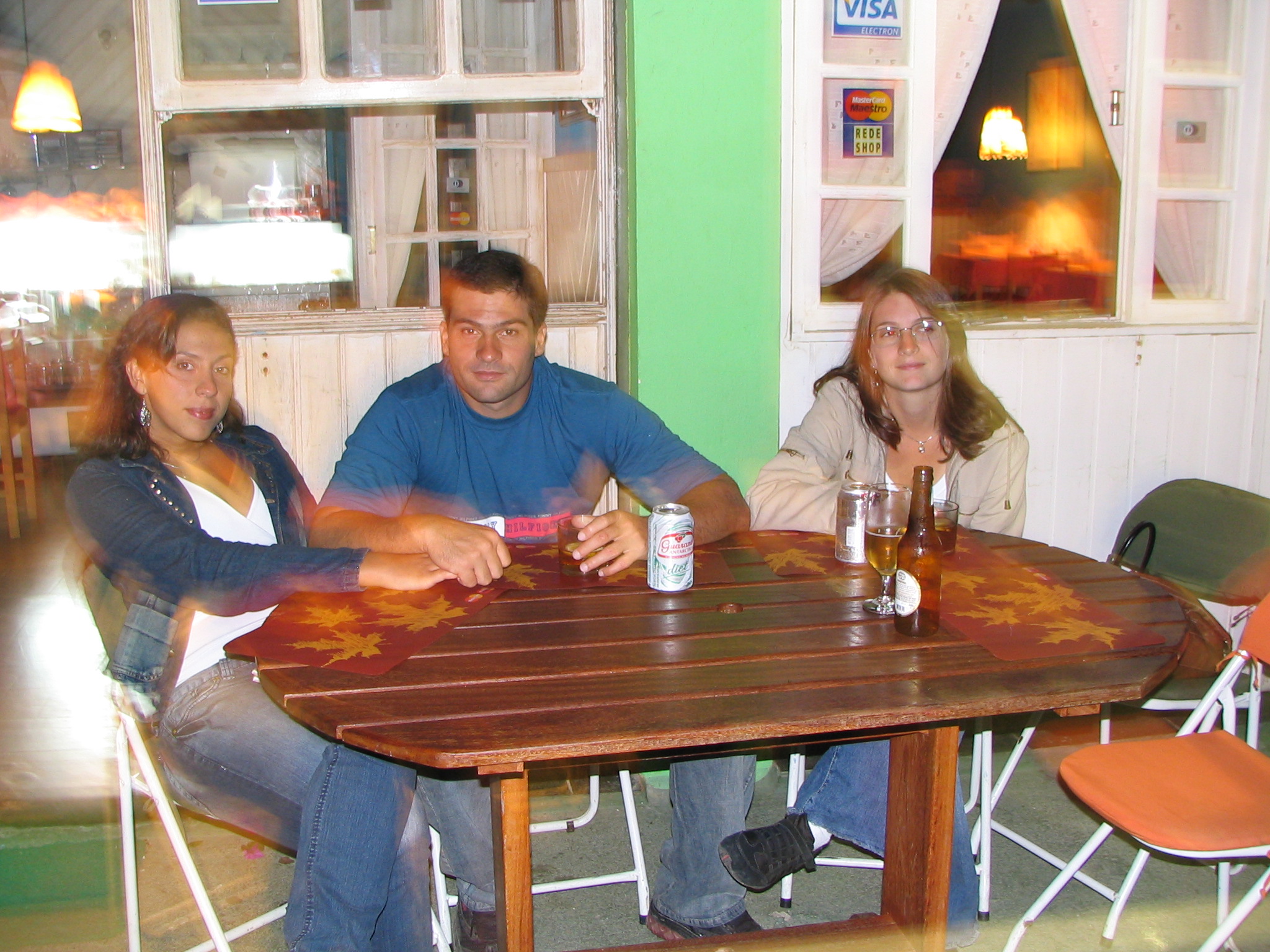}
		\caption{Failure case}
	\end{subfigure}
	\caption{(a) SFA-PLSR scores and \(2\sigma\)-confidence band on BID, (b) FRIQUEE scores and \(2\sigma\)-confidence band on BID, and (c) a failure case.}\label{fig:Failure}
\end{figure}

In this part, we further consider the prediction consistency of the proposed method and FRIQUEE (the method among the compared methods with the best overall performance). The green regions shown in Figure~\ref{fig:Failure}(a), (b) are the $2\sigma$-confidence bands on BID. The scatter points outside the band are regarded as outliers. It can be seen that FRIQUEE has more outliers than our method SFA-PLSR. The median values of outlier's ratio (OR) in 1000 runs are 5.98\%, 11.11\% for SFA-PLSR, FRIQUEE, respectively, which indicates that our method is more consistent with human perception. The outliers correspond the failure cases, and the worst case of our method is shown in Figure~\ref{fig:Failure}(c). The picture suffered from so complex distortions. To overcome this type of failure cases, more clues should be considered, such as saturation and ghosting.

\section{Conclusion}
\label{sec:conclusion}
In this paper, we propose a novel NR-IQA method for realistic blur images, which is based on statistically aggregating the high-level semantic features extracted from pre-trained deep convolutional neural networks. The top performance and strong generalization capability of our method are validated by comparing with several state-of-the-art methods on two realistic image databases (BID, CLIVE) and two synthetic image databases (TID2008, LIVE). Experiments also show that high-level semantics indeed play a more critical role than low-level features in NR-IQA of realistic blur images. In the future study, we will consider our methods in a coarse to fine multi-scale framework, since object scale also plays a role in human blur perception.

\begin{acks}
This work was partially supported by National Basic Research Program of China (973 Program) under contract 2015CB351803, the National Natural Science Foundation of China under contracts 61210005, 61390514, 61421062, 61527804, 61572042, 61520106004 and Sino-German Center (GZ 1025).
\end{acks}

\bibliographystyle{ACM-Reference-Format}
\bibliography{SFAforarxiv} 

\end{document}